# Radiation Damage Monitoring System of The CMS HF Detector

Yalcin Guler, [1]

[1]Department of Physics, Cukurova University, Institute of Science 01330, Adana, Turkey

Correspondence should be addressed to Y. Guler; yalcin.guler@cern.ch

## Abstract

Hadronic Forward(HF) detector is one of the sub-detectors of the CMS. It plays an important role in the measurement of forward jets and missing energy. HF is situated on both sides of the CMS interaction point, at about ±11 m, covers the very forward angles of CMS, in the pseudorapidity range ($3.0 < |\eta| < 5.0$), leaving only clearance for the beam pipes. This calorimeter consists of quartz fibers embedded in the iron absorber. Cherenkov light created by the charged particles is transmitted to PMTs by the fibers. Fibers due to the proton-proton collisions are exposed to high radiation. This radiation causes deterioration of the fibers over the time. Special channels are placed in the HF detector to observe the changes in the fibers due to radiation. It is possible to monitor radiation damage online using the data taken for these channels. In this study, radiation damage of HF is analyzed using the local data which was taken during 2010 run period. A number of innovations and modifications have been proposed for the analysis method.

## Introduction

Particle physics investigates what matter is made of at the most fundamental level. In order to understand this structure, scientists have put forward a theory called the Standard Model (SM) [1]. According to the SM, all matter in the universe is composed of point particles called quarks and leptons. The SM, which creates the matter and can explain the relationship between them, can answer many questions as a result of the experiments, but there are also some questions that it cannot answer. For this reason, scientists have turned to new theories beyond SM. Experimental physicists carry out researches at European Organization for Nuclear Research (CERN) to test these theories. The Large Electron-Positron Collider, which was built in a 27 km tunnel 100 meters below the ground in 1989, was dismantled and replaced by the Large Hadron Collider (LHC) in 2001 to collide hadrons.
At the LHC, p-p and ions collide in the previously unattained center of mass energy and luminosity. There are four major experiments on the LHC [2-5]. The CMS experiment is one of them. As a result of collisions using the high level of luminosity provided by the LHC, the forward parts of the detector are exposed to a high rate of radiation over time. The HF is exposed to the highest amount of radiation compared to other parts of CMS.

# Materials and Methods

## Materials

**Hadron Forward Calorimeter:** The HF shown in Figure 1 is located at 11.2 m from the interaction point and is made of steel absorbers and radiation hard quartz fibers [5-9]. This design lets HF a fast collection of the emitted Cherenkov light within the fibers due to charged relativistic particles. There are two sets of fibres: Long (1.65 m) and short (1.43 m) quartz fibers placed separately and read with separate phototubes. Using two different length of fibers let HF to discriminate Electromagnetic showers from Hadronic showers. Each HF module has 18 wedges in a non-projective geometry, with quartz fibers running parallel to the beam axis along the length of the iron absorbers [11-17]. The detectors are divided into a tower geometry. There are 13 towers in $\eta$, all with a size given by $\Delta\eta \approx 0.175$, except for the lowest-$\eta$ towers with $\Delta\eta \approx 0.1$ and the highest-$\eta$ towers with $\Delta\eta \approx 0.3$. The segmentation of all towers is 10°, except for the highest-$\eta$ which has $\Delta = 20°$. This leads to 900 towers and 1800 channels in the two HF modules [7].

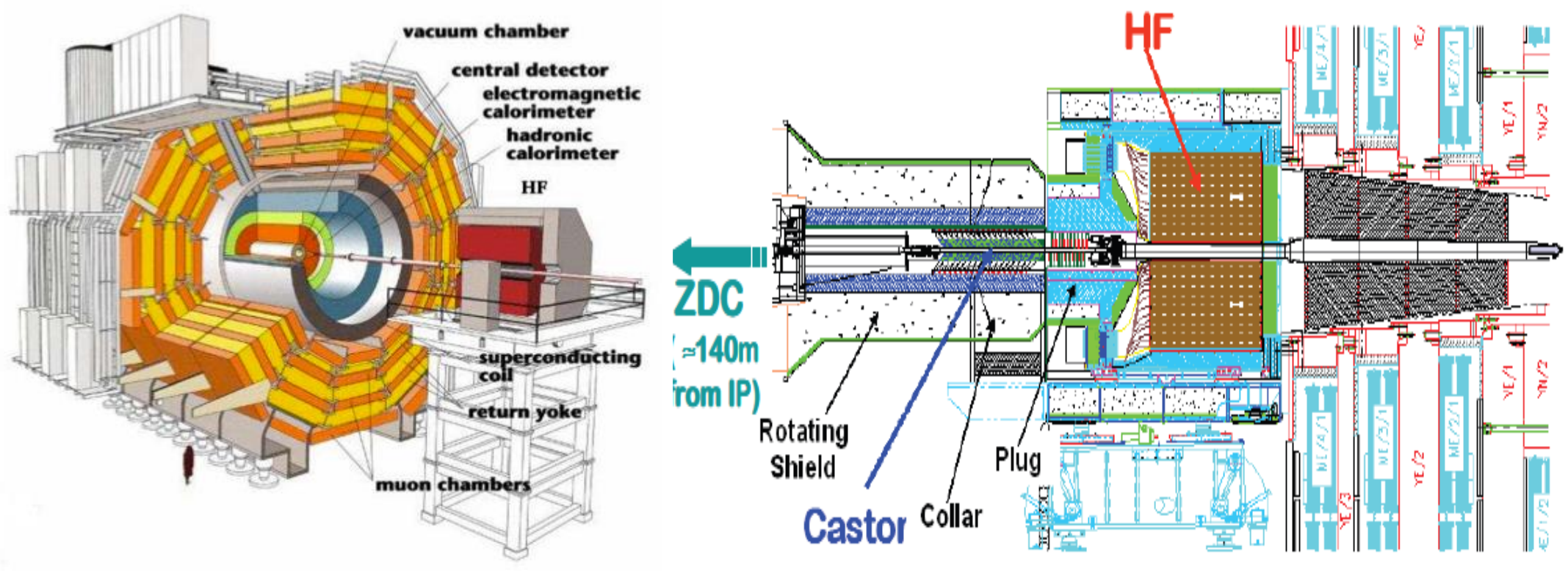

Figure 1. Overview of the CMS detector (left). The forward region on one side of the IP in CMS: the quartz calorimeter HF (right) [5].

## Radiation Damage Monitoring System in HF

Radiation Damage Monitoring System (RadDam) in HF is designed to monitor online the damage of the detector due to radiation [10]. Since different parts of the detector will be exposed to different radiation levels, a total of 56 RadDam fibers were placed in different towers of the detector. The working principle of the system is based on the comparison of the light pulse reflected from the input part of the detector with the light pulse reflected from the end of the detector [7,8].

## RadDam Setup in HF

The system used to monitor the effect of radiation on fibers is shown in Figure 2. In the system, a nitrogen laser that produces 337nm wavelength light pulses is used as the light source. Since this wavelength corresponds to the most absorbed wavelengths in the fiber, it is shifted towards the blue region (440 nm) thanks to the 2 cm long scintillation fiber placed in the feedback connector. This light is split into 4 wedges for each HF (Figure 2.). These sampling fibers were placed in seven towers, covering different regions within each wedge. Each fiber is read by the corresponding tower's PMT where the fiber is placed. A total of 56 sampling fibers are used to determine the damage caused by radiation [5,6].

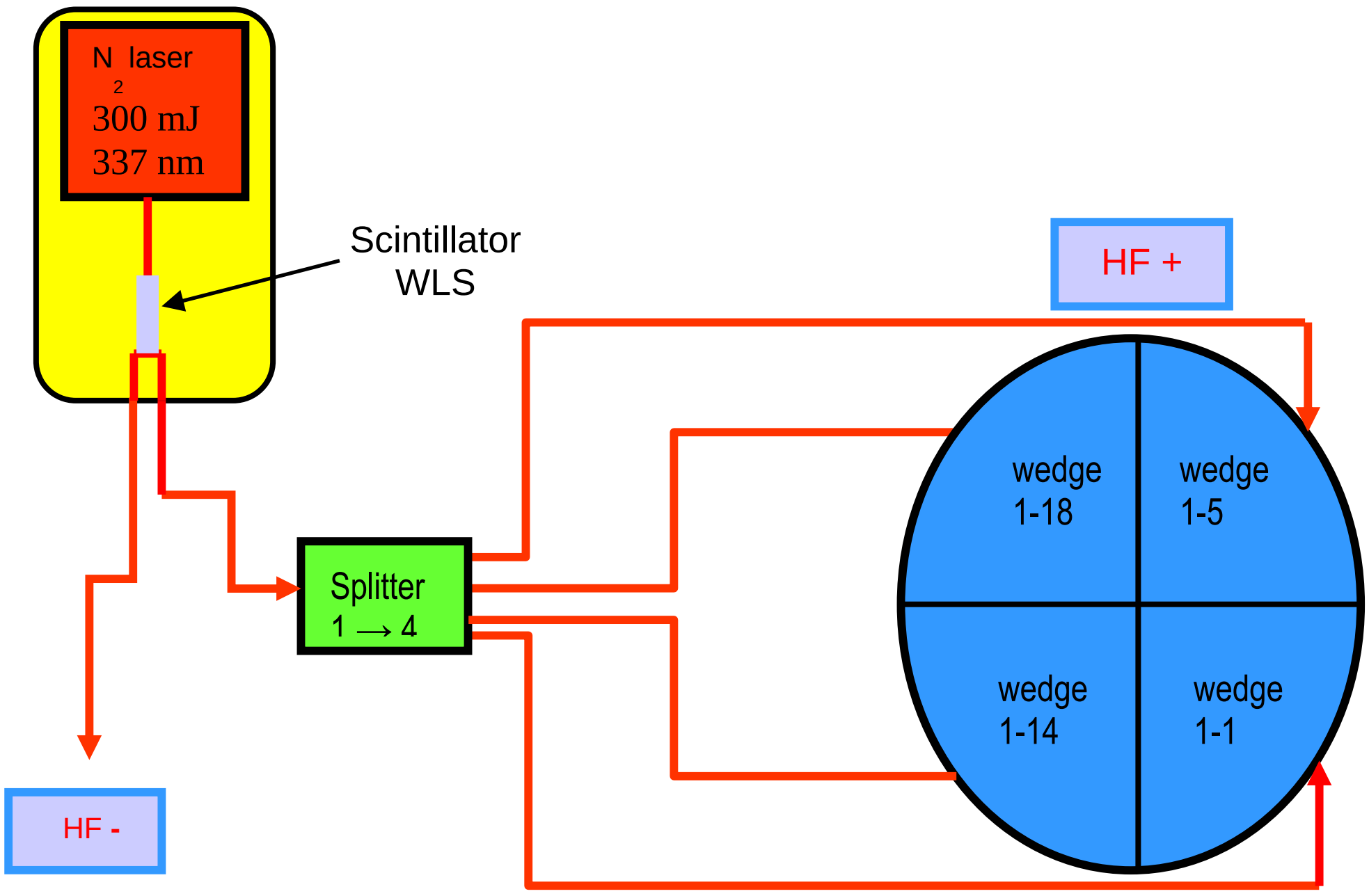

Figure 2. RadDam System. The light from the laser is shifted to a wavelength of 440 nm and distributed into four wedges through a splitter.

## RadDam Principle

There is some air gap in the junction area of the fiber coming from the laser and the fiber that is placed in the tower, as shown in Figure 3. In this way, a light pulse sent from the laser is reflected from the front of the fiber in the tower. The light transmitted to the fiber in the tower travels on this fiber and when it meets the air at the end of the fiber, another reflection occurs. Since the region of the first reflection occurs at the front of detector which is outside the detector, this reflection should not be affected by the radiation. However, since the second reflection is transmitted through the fiber in the tower, the amplitude of this reflection is expected to be attenuated due to the radiation. Therefore, the ratio of these two reflections shows us how much the fibers are affected by the radiation [17,18]. Two reflected signals are transmitted to PMT to be measured by another fiber. The second signal is detected by the photomultiplier tube after 25 ns, as it traverses the fiber in the tower twice. The reading system of the CMS is based on 10 consecutive time frames of 25 ns. Therefore, these two signals are located in two adjacent time slices (TS) (Figure 3). Fiber optic cables, which will be exposed to radiation as a result of long operation periods of the detector, will darken over time and will reflect less light. The decrease in light transmission will indicate how much the detector will be affected by the radiation over time. The RadDam channels mentioned above are shown with red boxes in the towers of the HF detector in Figure 4.

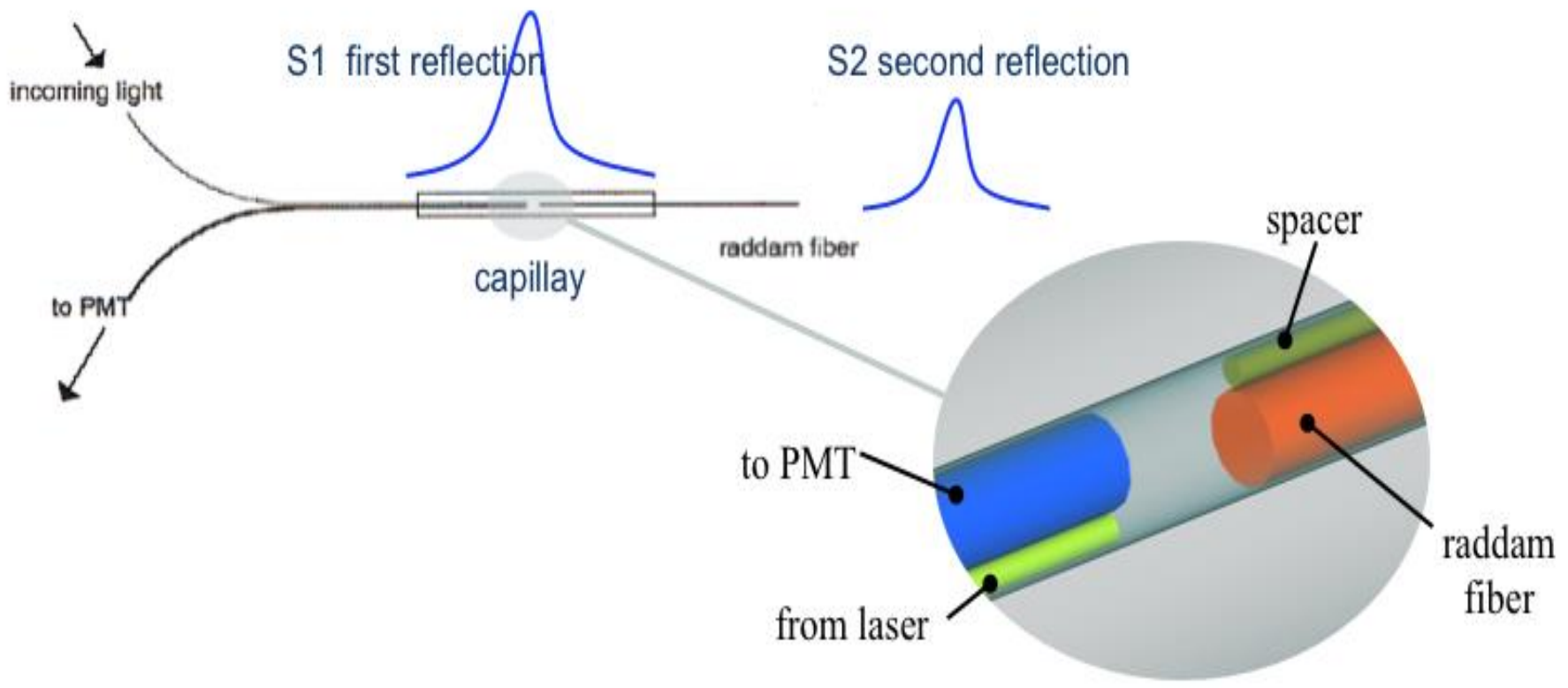

Figure 3. Representative picture showing the light from the laser traveling to the fiber in the tower and the reflected light being read by the PMT, as well as the shapes of the first and second reflected signals [10].

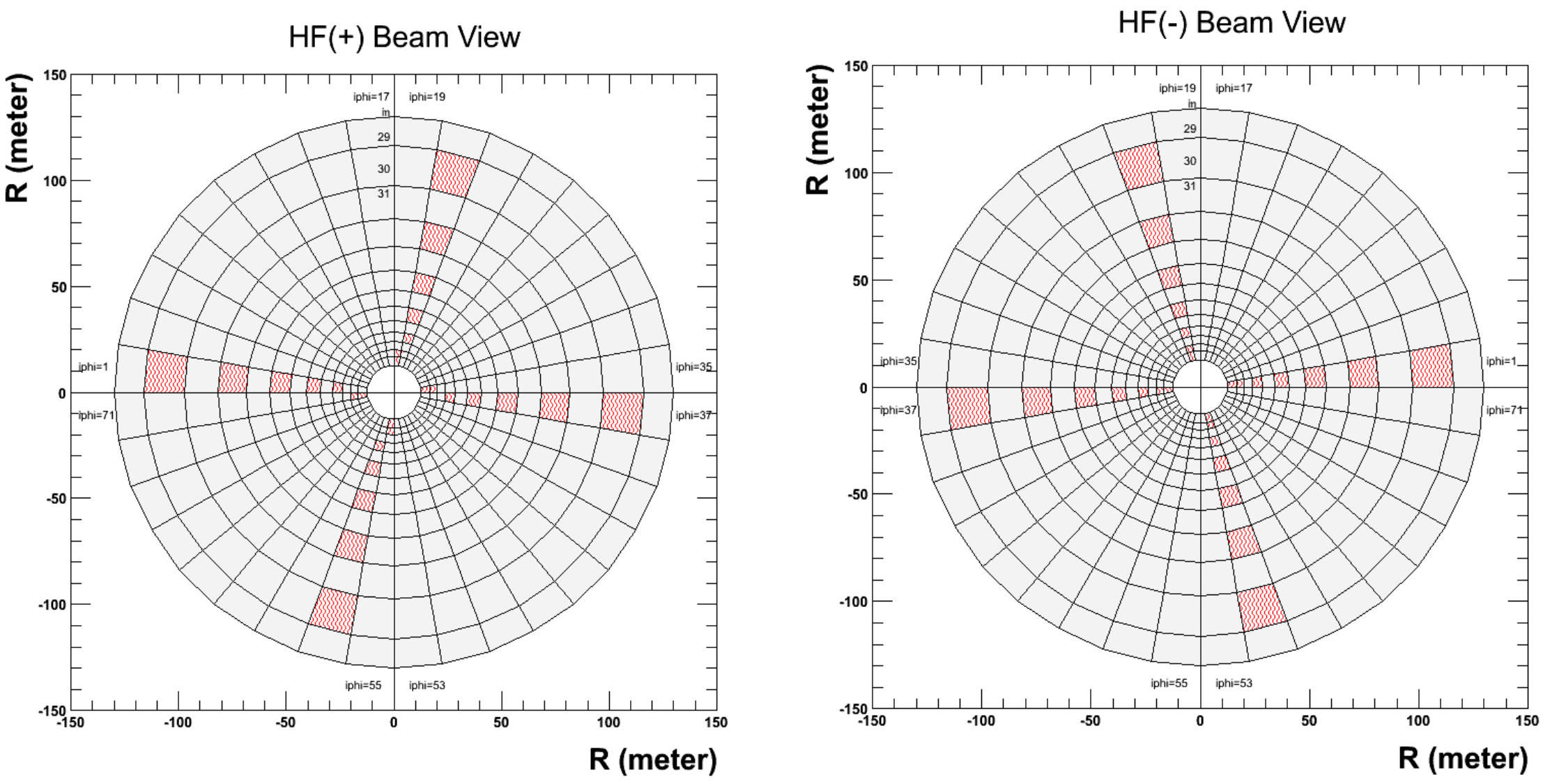

Figure 4. The locations of RadDam fibers are shown with red boxes.

## Method

As mentioned in the previous section, the system is based on the comparison of the ratios of two consecutive signals. In this section, the method used during the analysis of local data collected in 2010 will be given.

First of all, it is necessary to understand at which TS the signals obtained from the light reflection from the fibers in the received data are found. The histogram in Figure 5 shows the TS of the first and second reflected signals for only two channels.

## Event Selection

As can be seen from Figure 5, the first and second reflected signals for HF (+) are mostly in the 3rd and 4th TSs (a) and for HF (-) mostly in the 4th and 5th TSs (b). As can be seen from the figure, the signals are shared with the TSs in the neighborhood of these two TSs due to the shift of the laser pulse over time. In order for the ratio to be calculated accurately, the events where the signals are in two consecutive time slices should be selected.

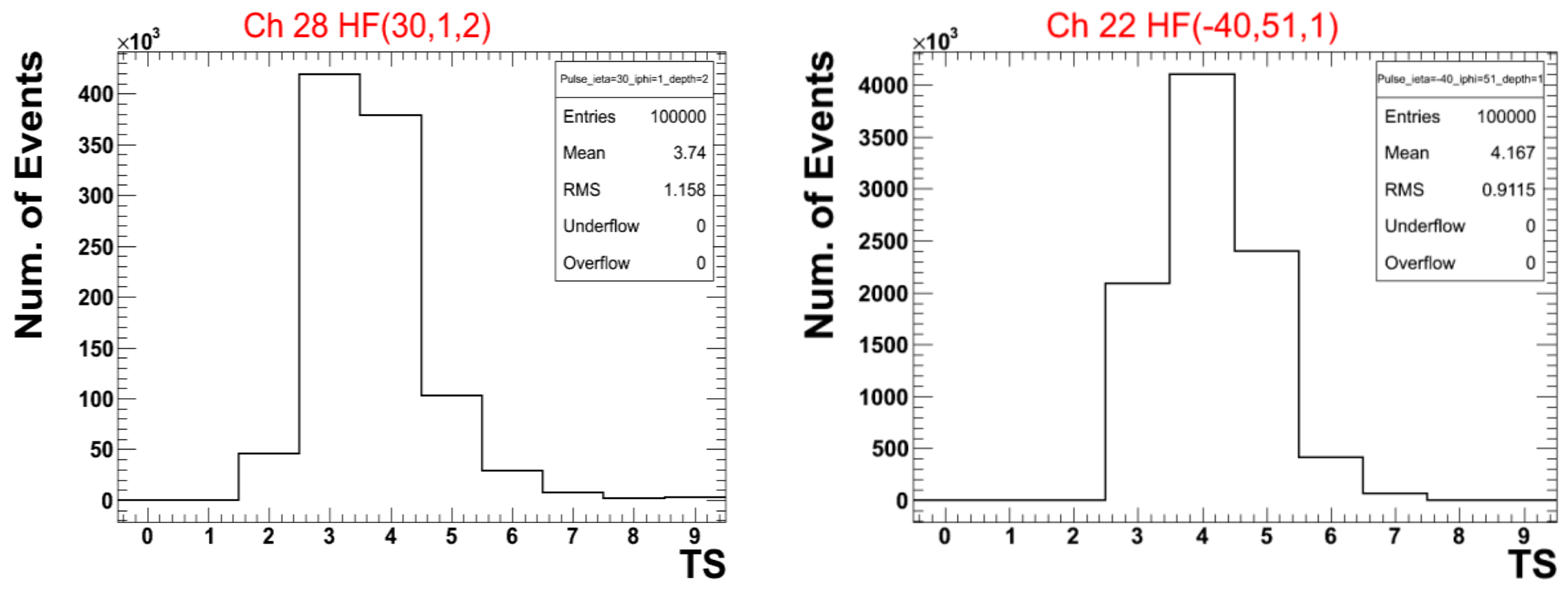

(a) (b)

Figure 5. a) Histograms are showing that the first and second reflected signal pulse shape for HF (+) is in the 3rd and 4th TSs, b) for HF (-) in the 4th and 5th TSs.

In this study, two selection cuts are applied to events in which the signals take place in two TS. The first condition is that the first signal (S1) and the second signal (S2) is greater than 5 fC. This condition is necessary to ensure that the signal is above the noise caused by the detector and to exclude some problems that may arise from the laser (for example, the fluctuation or shift of the pulses over time).
Figure 6. show charge distributions of S1 and S2, respectively. In these distributions, the *x*-axis shows the charge created by the signal, and the y-axis shows the number of events. It can be seen from this distribution that below 5fC is noise. Considering this distribution, the first event selection cut is applied as follows:

$$TS[J] > 5fC \text{ and } TS[J+1] > 5fC \tag{1}$$

It is important to select the events where the signal is in two TS after the noise cleaning. For this reason, it is required that 95% of the charge should be contained in the expected two consequtive TSs. Implement of this selection is formulated in Equation 2. According to that, the ratio of the amount of charge in the 3rd and 4th TSs(2Q) to the charge amount in the 2nd, 3rd, 4th, and 5th TS(4Q) for HF(+) is calculated. As an example of these TSs are showns in Figure 7 a for one event for HF(+). The distribution of the above mentioned ratio is shown in Figure 7 b for all events. This selection cuts should eliminate events in which signals are not in two TSs. After these two selections, 20-30% of the total data remains.

$$\frac{TS[J]+TS[J+1]}{TS[J-1]+TS[J]+TS[J+1]+TS[J+2]} > 0.95 \tag{2}$$

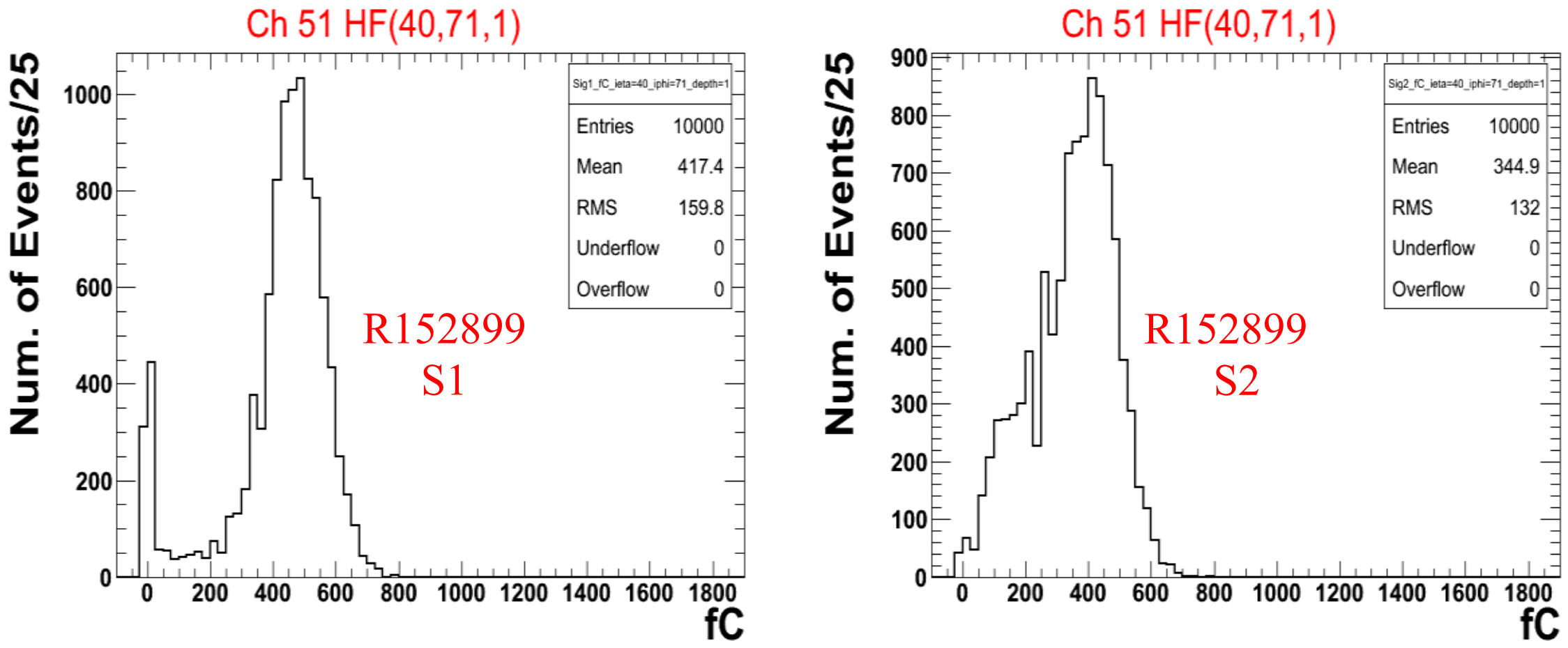

Figure 6. Charge distributions of the first and second signals.

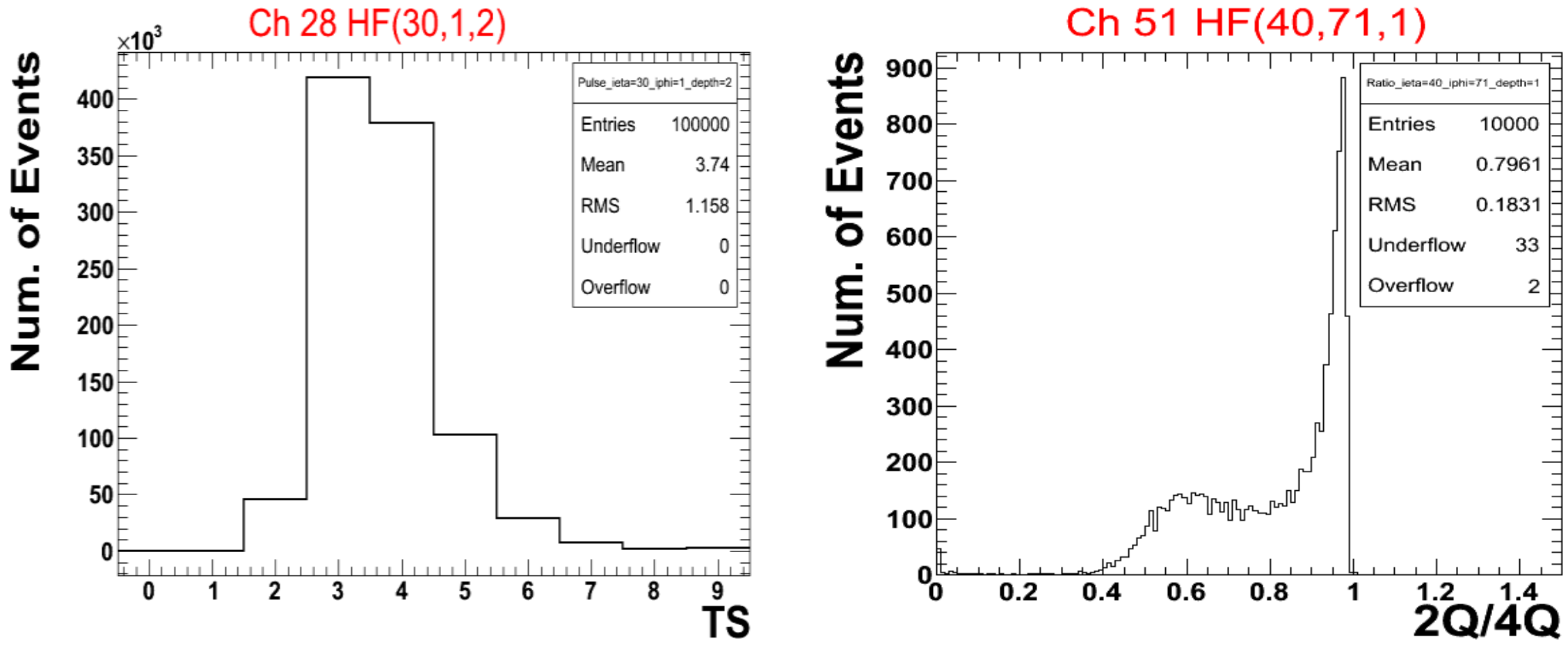

(a) (b)

Figure 7. Representation of the second condition in the selection of events. The histograms used to decide the event selection.

## Results and Discussion

The event selection described above were applied to data from 56 channels designed to monitor radiation damage. Except for a few problematic channels, the signal distribution of the other channels is similar to the distribution of the 51st channel shown in Figure 8.a, b. It can be seen that the average charge values of the first reflected and second reflected signals for each channel are not the same (Figure 9.a, b). This is a known behavior because not all channels are equally identical. The important point is to have enough signal height for this analysis. Since determining a fixed histogram range will cause an increase in statistical error in some channels, it is necessary to select the appropriate histogram range for the data from each channel. An important point to consider when creating histograms is to choose the histogram range in a way that does not change the arithmetic mean.

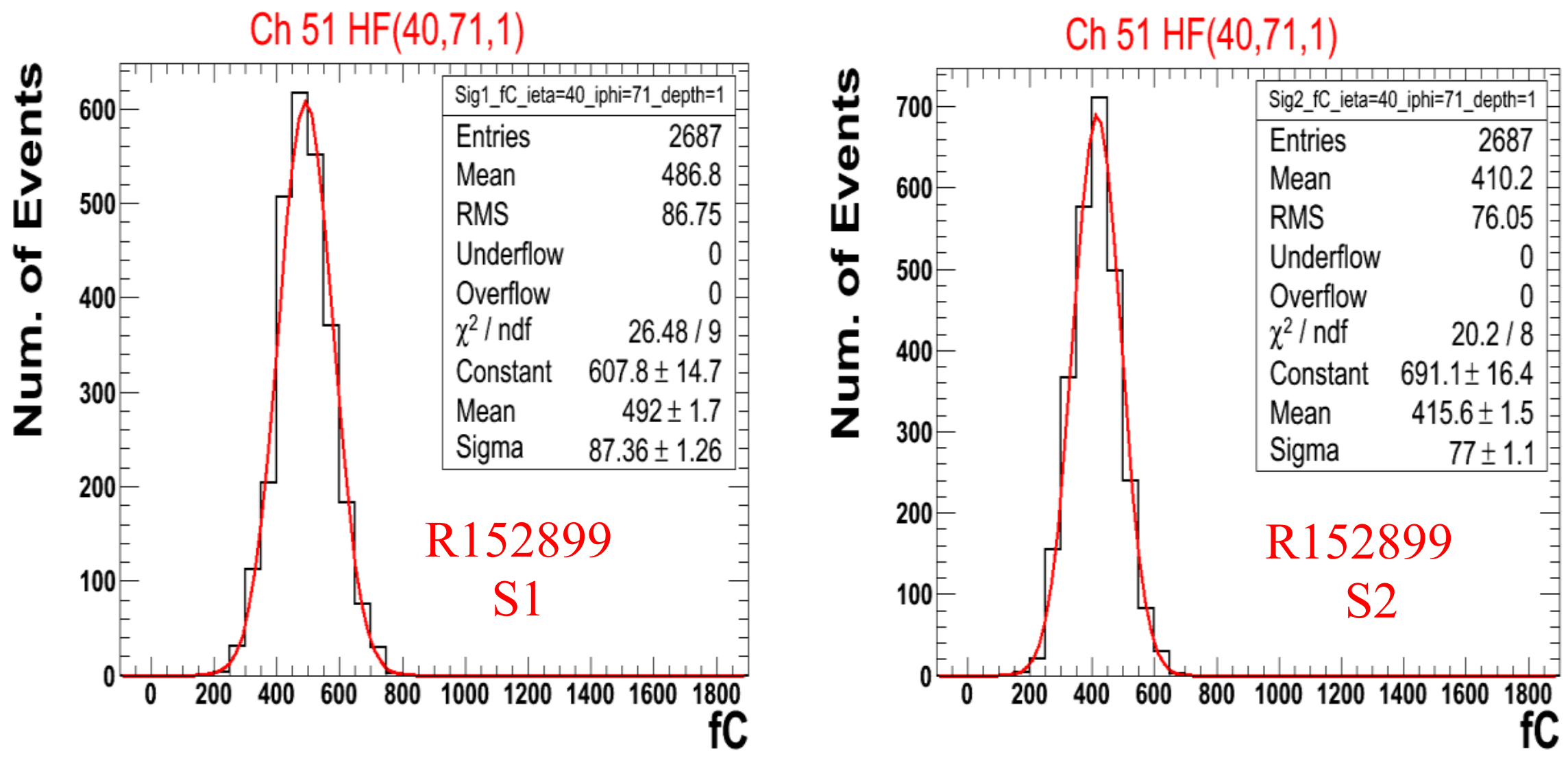

(a) (b)

Figure 8. First and second reflected signal distributions are fitted to Gaussian function.

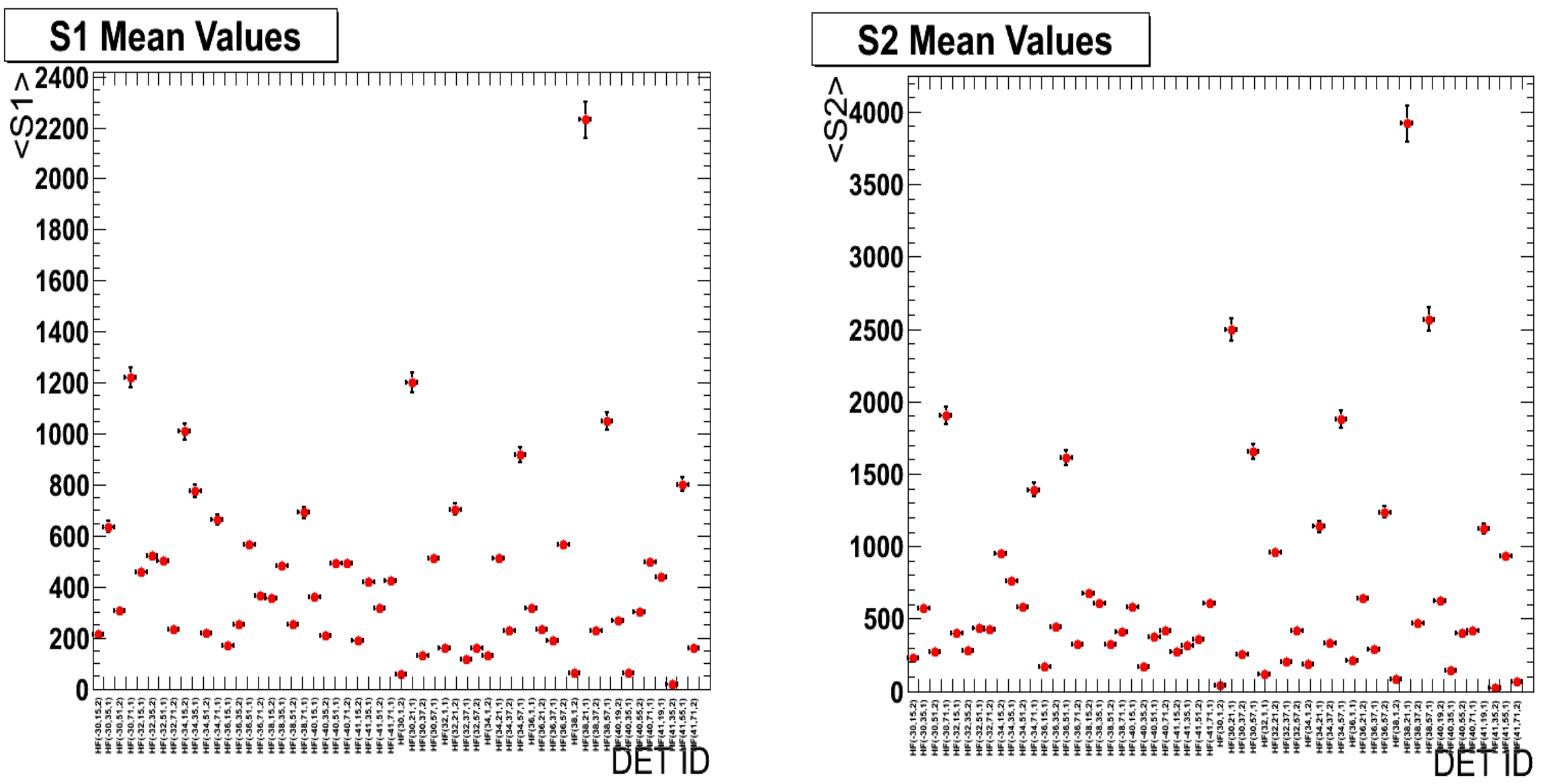

(a) (b)

Figure 9. Mean values of the first and second signals of each channel.

## Error Sources

In this study, the sources of erros can be divided in two: as errors coming from the method and from the system.

## Method Errors

In order to understand the errors coming from the method, it is necessary to compare the data obtained at different times. Thus, the stability of the method is checked with data taken at different times. It is expected that the results should not be time dependent. Two data sets taken at different times were analyzed separately by the method described above and the results from each channel were compared with the result of the same channel in the other data using the following equation. The results can be seen Figure 10. The average value of this histogram gives us the range of error of the method, which was found to be around 1%.

$$\frac{\Delta}{\text{mean}} = \frac{|R(t)_a - R(t)_b|}{0.5\times[R(t)_a + R(t)_b]} \quad (3)$$

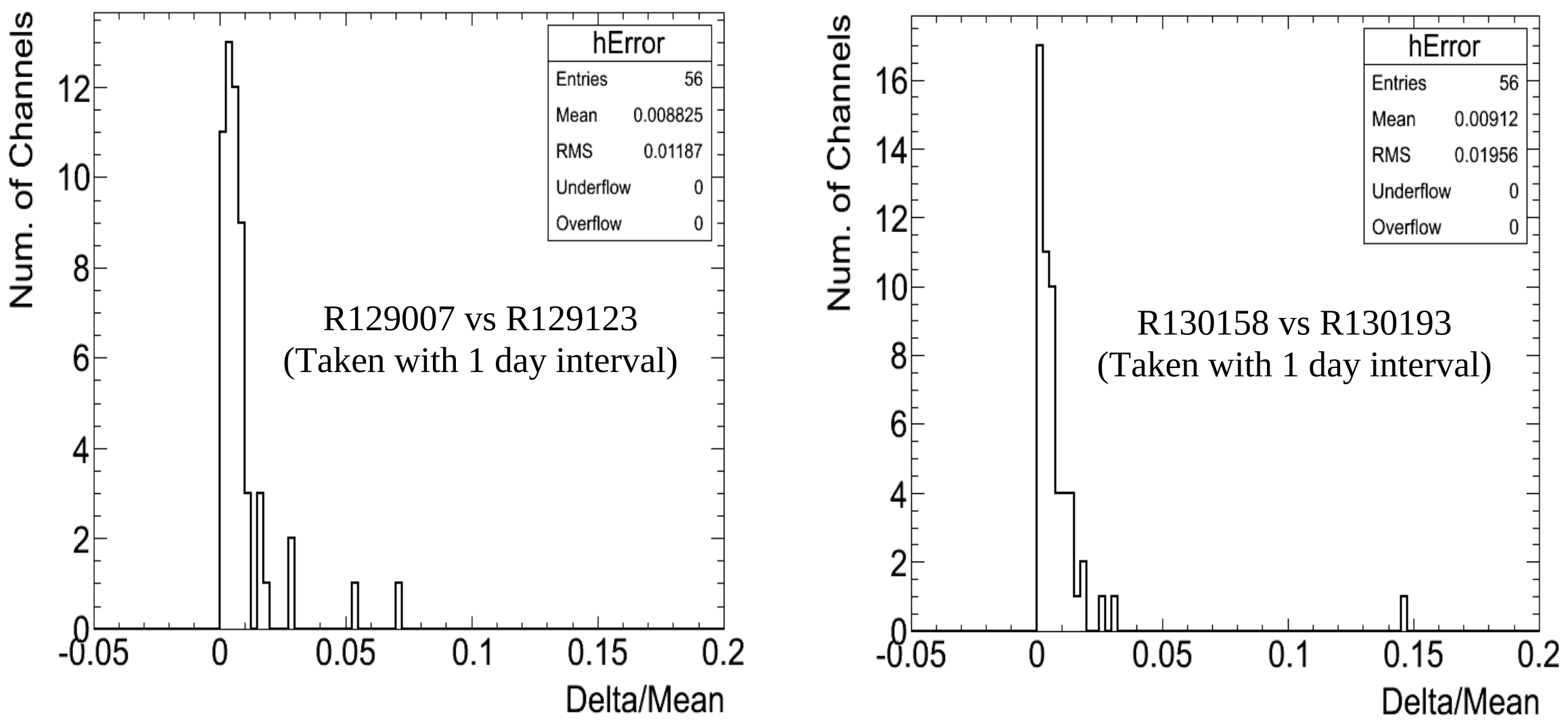

(a) (b)

Figure 10. Error rates found using the data taken one day apart.

**Systematic Error**

QIE cards are known to not always be able to divide the data correctly. For this reason, the error is % 2 [19]. The error caused by its sharing between QIEs is 2.5%.

We can combine the contribution from different sources by taking the square root of the sum of the squares of these errors; this means the error is 3.35%.

**Radiation effect**

In Figure 11, each point in the histogram shows the ratio of the mean values of the first reflected signal distribution of a channel to an initial (reference) data. The trend of this ratio has been observed for each channel with all the data received. Figure 12 shows the second signal in a similar way.

In order to determine the damage caused by radiation, the ratio given by Equation 4 should be calculated and plotted as a function of time. The obtained by this method is shown in Figure 13. The given result is only for one channel. The behaviour of the other channels is similar.

$$R_t/R_0 = [S_2(t)/S_1(t)]/[S_2(0)/S_1(0)] \quad (4)$$

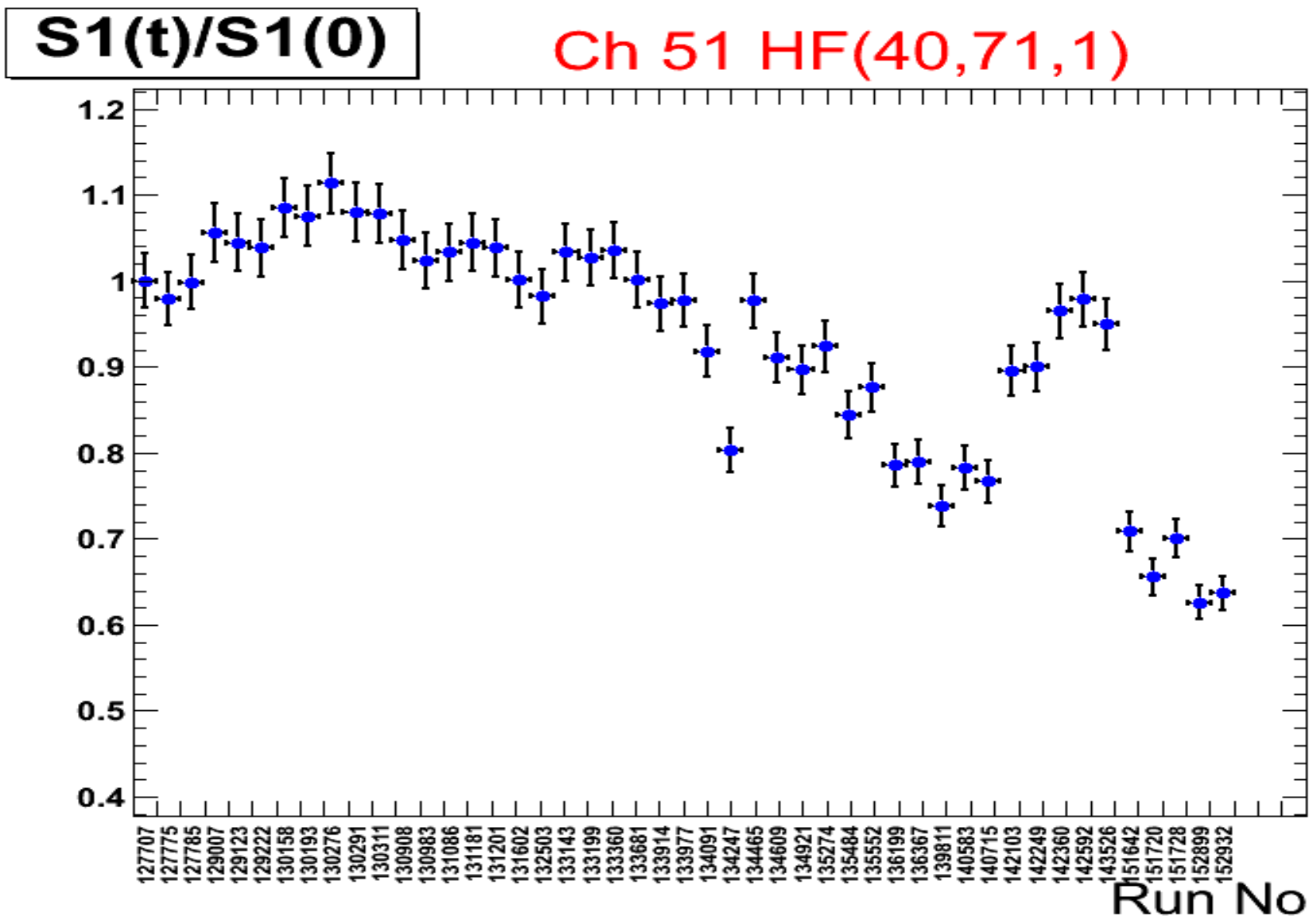

Figure 11. Histogram showing the behavior of the first signal over all data relative to a reference data, for only one channel.

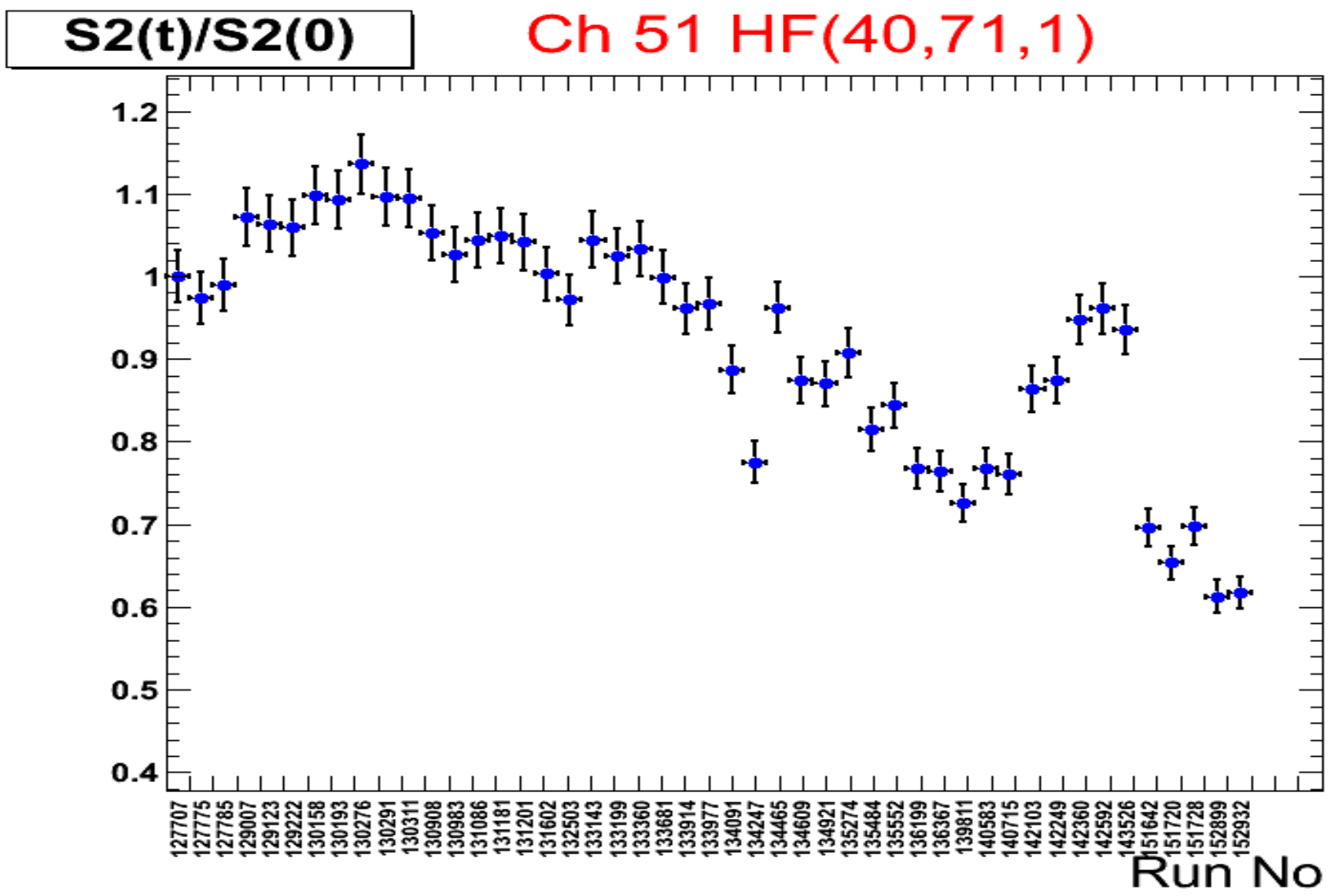

Figure 12. Histogram showing the behavior of the second signal over all data relative to a reference data, for only one channel.

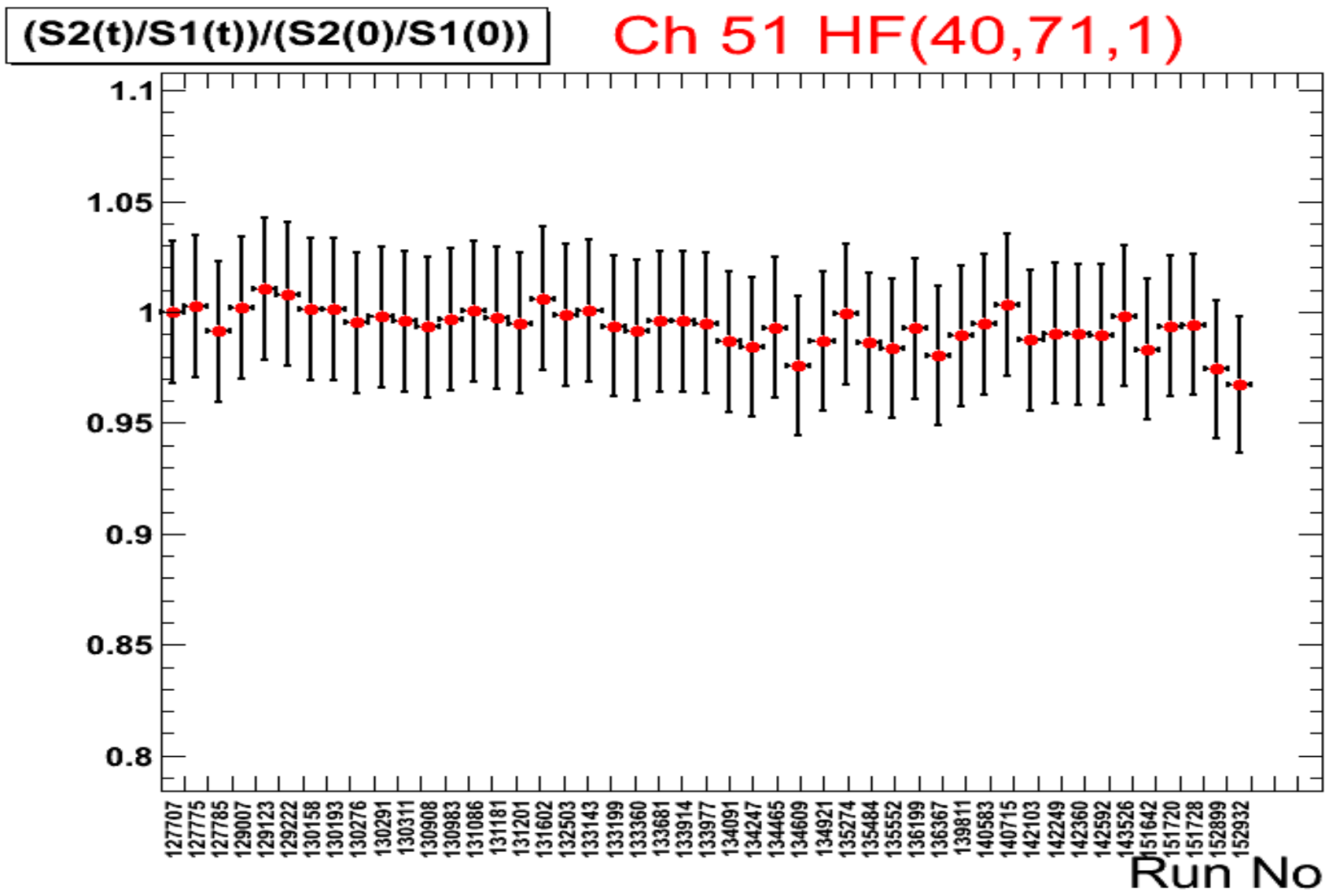

Figure 13. Histogram is shown the radiation damage of just one channel over all data.

## Conclusions

The hadronic calorimeter (HCAL) is one of the subsystems in CMS. HCAL is the detector devoted to the energy measurement of hadrons and the missing transverse energy of events [20,21]. Hadronic calorimeters are important detector in particle physics experiments. In order to be able to use them reliably, they need to be well-calibrated. Energy is calibrated using known particles (radioactive sources) or using LED and Laser light [22]. Due to various reasons (radiation exposure, electronics, individual channels responses, etc) channels will measure different energy than the truth . Thus, the measured energy will not be correct and some correction coefficient will be needed. HF is exposed to the most intense radiation of all CMS subdetectors. The amplitude of the light transmitted by the quartz fibers attenuates over time due to the radiation. In this case, energy calibrations and corrections should be applied [23,24].

In this study, radiation damage to the quartz fibers, which is the active material of the HF calorimeter in the forward region of CMS, was determined and analyzed. With the analysis of the data received from RadDam channels, a system that monitors the damage online was created and developed. As a result of the analysis of the laser signals received from 56 channels placed in 8 wedge and 7 towers of the HF detector, it was seen that the received signal in only one of the channels could not be properly analyzed because not enough data was available and there were different behaviors in two channels compared to the other channels.

Throughout 2010, both the development of the method and the observation of the radiation damage to the detector were carried out on 47 data sets received from the system. As of 2010, there was no observable radiation damage in the detector.